\documentclass[%
 amsmath,amssymb,
 aip,
reprint,
]{revtex4-1}

\usepackage{natbib}
\usepackage{url}
\usepackage{natbib}
\usepackage{multirow}
\usepackage{amssymb}
\usepackage{mathtools}
\usepackage{xcolor}
\usepackage{amsmath,pgfmath}
\usepackage{pifont}

\usepackage[symbol]{footmisc}

\begin{document}

\markboth{Author et al.}{Neural Network Potentials}

\title{Neural Network Potentials: A Concise Overview of Methods}


\author{Emir Kocer}
\thanks{These authors contributed  equally to this paper.}
\affiliation{Universit\"{a}t G\"{o}ttingen, Institut f\"{u}r Physikalische Chemie, Theoretische Chemie, Tammannstra\ss{}e 6, 37077 G\"{o}ttingen, Germany; email: ekocer@uni-goettingen.de; ORCID: 0000-0003-4207-8220}

\author{Tsz Wai Ko}
\thanks{These authors contributed  equally to this paper.} 
\affiliation{Universit\"{a}t G\"{o}ttingen, Institut f\"{u}r Physikalische Chemie, Theoretische Chemie, Tammannstra\ss{}e 6, 37077 G\"{o}ttingen, Germany; email: tko@uni-goettingen.de; ORCID: 0000-0002-0802-9559}

\author{J\"{o}rg Behler}
\thanks{Corresponding author: joerg.behler@uni-goettingen.de}%
\affiliation{Universit\"{a}t G\"{o}ttingen, Institut f\"{u}r Physikalische Chemie, Theoretische Chemie, Tammannstra\ss{}e 6, 37077 G\"{o}ttingen, Germany; email: joerg.behler@uni-goettingen.de; ORCID: 0000-0002-1220-1542}

\begin{abstract}
In the past two decades, machine learning potentials (MLP) have reached a level of maturity that now enables applications to large-scale atomistic simulations of a wide range of systems in chemistry, physics and materials science. Different machine learning algorithms have been used with great success in the construction of these MLPs. In this review, we discuss an important group of MLPs relying on artificial neural networks to establish a mapping from the atomic structure to the potential energy. In spite of this common feature, there are important conceptual differences, which concern the dimensionality of the systems, the inclusion of long-range electrostatic interactions, global phenomena like non-local charge transfer, and the type of descriptor used to represent the atomic structure, which can either be predefined or learnable. A concise overview is given along with a discussion of the open challenges in the field. 
\end{abstract}

\maketitle


\section{Introduction and Scope of this Review}
Machine learning (ML) plays an increasingly important role in many aspects of life~\cite{editors2018machine,alpaydin2020introduction}. The ability to provide, after a training process, accurate predictions make modern ML algorithms a very attractive tool for applications in all fields of science, in which the analysis, classification and interpretation of data is important~\cite{P5779,P6098,P6099,P6083,libbrecht2015machine,selvaratnam2021machine}. Consequently, data science and machine learning are now called the ``fourth paradigm of science''~\cite{P5651} along with the three established paradigms of empirical experimental studies, theoretical models, and computer simulations. 

A particularly fruitful combination of machine learning and computer simulations that has emerged in the past two decades is the representation of the multidimensional potential energy surface (PES) by machine learning potentials (MLP) \cite{P2559,P3033,P4885,P5788,P5673,P5793,P6037,P6101,P6102,P6112,P5928}. The PES is of central importance for reaching an atomic-level understanding of any type of system, from small molecules to bulk materials, as it contains all the information about the stable and metastable  structures, the atomic forces driving the dynamics at finite temperatures, the transition states and barriers governing reactions and structural transitions, and also the atomic vibrations. For a long time computer simulations in chemistry, molecular biology and materials science either had to rely on computationally demanding electronic structure calculations like density-functional theory (DFT) or efficient but significantly less accurate empirical potentials or force fields based on physical intuition and approximations. Modern MLPs bridge this gap by learning the shape of the PES from reference data obtained from high-level electronic structure calculations. The resulting analytic MLPs represent the atomic interactions, commonly as a function of the atomic positions and nuclear charges, and can then be used in large-scale simulations like molecular dynamics (MD) many orders of magnitude faster than the underlying electronic structure calculations without a significant loss in accuracy. MLPs thus represent an important tool to meet the continuously increasing need for computer simulations of more and more complex systems.

Since the advent of the first MLP in 1995~\cite{P0316}, many different types of methods have been proposed, like neural network potentials (NNP)~\cite{P0421,P1174,P5366,P5577}, Gaussian approximation potentials (GAP)~\cite{P2630,P4429}, kernel-based approaches like gradient domain ML (GDML)~\cite{P5684}, spectral neighbor analysis potentials (SNAP)~\cite{P4644,P5993}, moment tensor potentials (MTP)~\cite{P4862}, atomic cluster expansion (ACE)~\cite{P5794}, atomic permutationally invariant polynomials (aPIP)~\cite{P5937} and support vector machines (SVM)~\cite{P4479}. MLPs offer many advantages, like a very flexible functional form that allows to represent the available reference data with a very high accuracy~\cite{P0422}, generality making them suitable for all types of bonding and atomic interactions, from covalent bonds via metallic bonding to dispersion interactions, and a well-defined analytic form enabling the consistent computation of energies and gradient-based properties like forces and the stress tensor. The main disadvantage of MLPs is the lack of a physical functional form and the resulting limited transferability beyond the structural diversity of the underlying training data, making a careful validation of the obtained potentials an essential step. For the same reason, the construction of MLPs is computationally very demanding, since large training sets are required to ensure that all required information about the topology of the PES can be learned from the reference data. 

The applicability of current MLPs is very diverse in many respects, e.g., concerning the dimensionality of the systems, the range of the atomic interactions that can be described, and the possible inclusion of physical laws and concepts. Recently, a classification scheme for MLPs into four generations has been proposed~\citep{P5977,p6018,P5932}, which we will also adopt in this review. First-generation MLPs can provide accurate PESs for low-dimensional systems like small molecules. Second-generation MLPs make use of the locality of a major part of the atomic interactions and employ the concept of atomic energies making these potentials applicable to high-dimensional systems containing thousands of atoms. Third-generation MLPs overcome this locality approximation by explicitly including in addition long-range interactions like electrostatics based on Coulomb's law without truncation. Still, they rely on atomic charges depending on the local chemical environment. Finally, fourth-generation potentials are able to capture phenomena like non-local charge transfer by taking into account the global structure and charge distribution of the system. 

In this review, we will focus on discussing MLPs based on neural networks (NN). NNPs do not only have the longest history, but also by far the largest diversity in terms of methods and concepts of all classes of MLPs. We will restrict our discussion to methods providing continuous PESs for arbitrary atomic positions suitable for molecular dynamics simulations, while we note that many other interesting NN-based approaches have been proposed and successfully tested to predict, e.g., atomization energies of large sets of organic molecules in their global minimum geometries \cite{P3136} and a variety of properties of materials~\cite{zeng2018graph,omprakash2021graph,miccio2020chemical}. Further, we do not include methods used for predicting atomic partial charges or multipoles unless they are used to explicitly compute the electrostatic energy contribution to the PES. Our aim is to present the main methods and concepts, while applications are only briefly mentioned. The interested reader is referred to many reviews providing comprehensive lists of such applications~\cite{P2559,P3033,P5128,P5793}. 

\begin{figure*} 
    \centering
    \includegraphics[width=14cm]{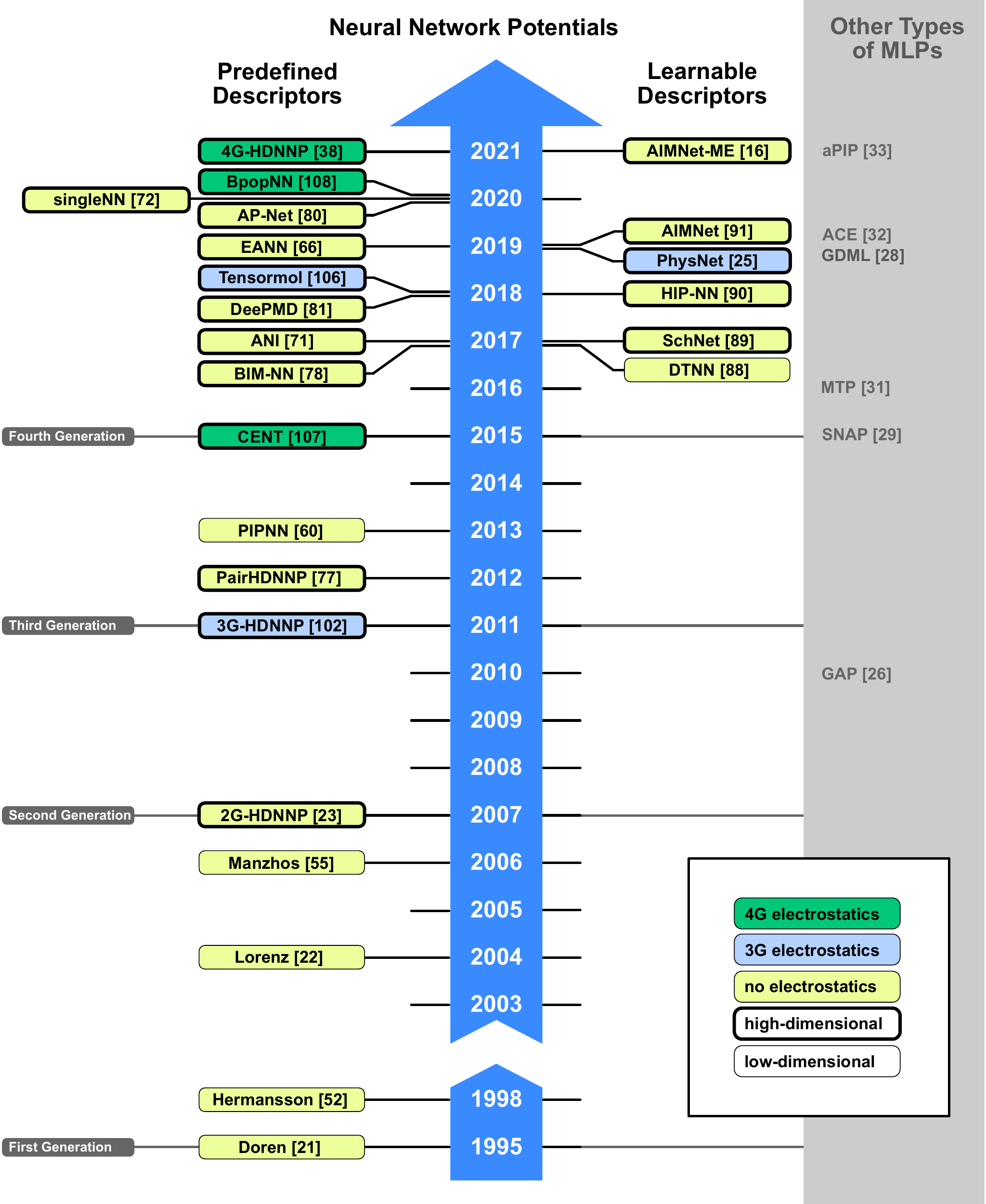}
    \caption{Timeline for the evolution of neural network potentials. NNPs using predefined descriptors are shown on the left, NNPs with learnable descriptors (message passing neural networks) on the right. The frame width and background color label the dimensionality and the absence or inclusion of long-range electrostatic interactions in third (3G) and fourth (4G) generation potentials. Some other important types of MLPs not discussed in this review are given with references in grey on the right.}
    \label{fig:timeline}
\end{figure*}

Even with this restricted scope, giving a comprehensive overview about all available methods in this rapidly expanding field is an  impossible endeavor. Figure~\ref{fig:timeline} shows a schematic timeline of the historic evolution of NNPs using representative examples. The presented methods are classified based on the types of descriptors of the atomic configurations, which can be predefined or learnable, the dimensionality of the systems that can be studied, and the inclusion or absence of long-range electrostatic interactions. These criteria will provide the central guidance for the discussion of the individual methods in this review in the following sections. In addition, the evolution of NNPs is put into the context of other important MLP methods proposed to date, which are given in the right part of Fig.~\ref{fig:timeline} but are not discussed here in more detail.

\section{The First Generation: Low-Dimensional Systems}

In 1995, Doren and coworkers published the first machine learning potential trained to electronic structure data enabling simulations at strongly reduced costs~\citep{P0316}. They used a feed-forward neural network to construct a mapping from the atomic positions to the potential energy for the adsorption of H$_2$ on a cluster model of the Si(100) surface. In this pioneering work, many technical problems have been investigated and solved, which nowadays are considered as textbook knowledge about MLPs, and the advantages and limitations of machine learning for representing PESs have been thoroughly discussed. In the following decade, the applicability of neural networks has further been explored by several groups for numerous molecular systems~\cite{P0420,P0833,P0840,P2394}. Another branch of applications addressed adsorption and gas-surface dynamics~\cite{P0421,P1820,P3371,P3152}. The construction of NNPs for condensed systems like liquid and bulk materials was out of reach at that time, but several attempts have been made to augment empirical potentials suitable for larger systems by neural networks~\cite{P0838,P0830,P0837}.

In addition to these applications, very promising methodical advances have been made, resulting e.g. in very accurate potentials based on a sum of products, optimized coordinates and a high-dimensional model representation proposed by Manzhos and Carrington Jr., which reached spectroscopic accuracy \cite{P0924,P1390,P2059}. In 2009, Malshe et al. proposed to use a set of neural networks for the different terms in a many-body expansion~\cite{P2210}. Finally, in 2013 Jiang, Li and Guo used permutationally invariant polynomials (PIP), originally developed by Braams, Bowman and coworkers for the construction of PESs with full permutation invariance \cite{P3062,P2881}, as input coordinates for constructing molecular PIP-NNs~\cite{P3873,P3973}.

In spite of the diversity of methods and systems that have been studied in the initial decade, the common feature of all these first-generation MLPs, which exclusively relied on neural networks, is the small number of degrees of freedom that could be taken into account explicitly, limiting applications to small molecules in vacuum, molecules interacting with frozen surfaces, or the selection of specific low-dimensional additive terms, often in combination with conventional empirical potentials or force fields. A summary of these applications can be found in two early reviews~\cite{P2559,P3033}. 

\section{The Second Generation: Local Methods for High-Dimensional Systems}
\subsection{Overview}

Many important advances have been made in first-generation NNPs, but the use of first-generation NNPs remained limited to rather low-dimensional systems, which include the degrees of freedom of a few atoms only. The main conceptual problem preventing the application of MLPs to high-dimensional systems containing thousands of atoms has been the lack of suitable structural descriptors including the mandatory invariances of the total energy with respect to translation, rotation and permutation, i.e. the order, of chemically equivalent atoms in the system. Further, the use of a single neural network to express the global energy of the system, which is a central component of many early NNPs, did not allow the application to systems containing variable numbers of atoms, as the dimensionality of the system is related to the number of input neurons, which cannot be changed after the training of the NN.

Many empirical potentials in the field of materials science construct the total energy as a sum of atomic energies,
\\
\begin{eqnarray}
E_{\mathrm{total}}=\sum_{i=1}^{N_{\mathrm{atoms}}} E_i \quad, \label{eq:2Getot}
\end{eqnarray}
\\
and also most classical force fields used, e.g., in biomolecular chemistry can be cast into this form. For these types of potentials incorporating all invariances into the potential is straightforward, since they consist of very simple low-dimensional terms, which are additive and rely on translationally and rotationally invariant internal coordinates like interatomic distances, angles, and dihedral angles. This is different in case of MLPs, which have a more complex multidimensional functional form allowing to reach a very high numerical accuracy and require an ordered vector of input coordinates lacking intrinsic permutation invariance. Consequently, in the first decade of MLP development it has not been possible to make use of Eq.~\ref{eq:2Getot}, and the search for suitable descriptors taking all invariances into account has been a frustrating challenge with only a few special-purpose solutions for low-dimensional systems~\cite{P0830,P1388}. 

This problem has now been solved, and from today's perspective it is hard to imagine the difficult situation of early MLP developers, because a huge variety of different descriptors for high-dimensional systems is now readily available \cite{P5803,P5658}. The resulting very powerful second generation of MLPs based on Eq.~\ref{eq:2Getot} is applicable to systems of in principle arbitrary size, and many combinations of the two main ingredients of modern MLPs, the descriptors of the atomic environments and the ML algorithm connecting the descriptor values to energies, have been explored. 

Two major classes of second-generation NNPs can be distinguished depending on the type of descriptor. The first high-dimensional NNPs that became available made use of descriptors with predefined functional forms, which contain a few parameters defining the spatial shape. In this respect, they share some similarities with basis functions used in electronic structure calculations. Sometimes these parameters are even optimized during the fitting process \cite{P5292,P5792}. Due to their simplicity and efficiency, such predefined descriptors are still dominantly used in applications of current MLPs. In addition, in recent years a second class of second-generation NNPs has received increasing attention, which is based on the automatic learning of descriptors from structural information using message passing neural networks. This is not to be confused with the optimization of the parameters of the predefined descriptors, since no ad-hoc assumptions about the functional form are made. Both approaches
will be presented in the subsequent sections and the most important examples will be briefly discussed. 
    
\begin{figure*}
    \centering
    \includegraphics[width=14cm]{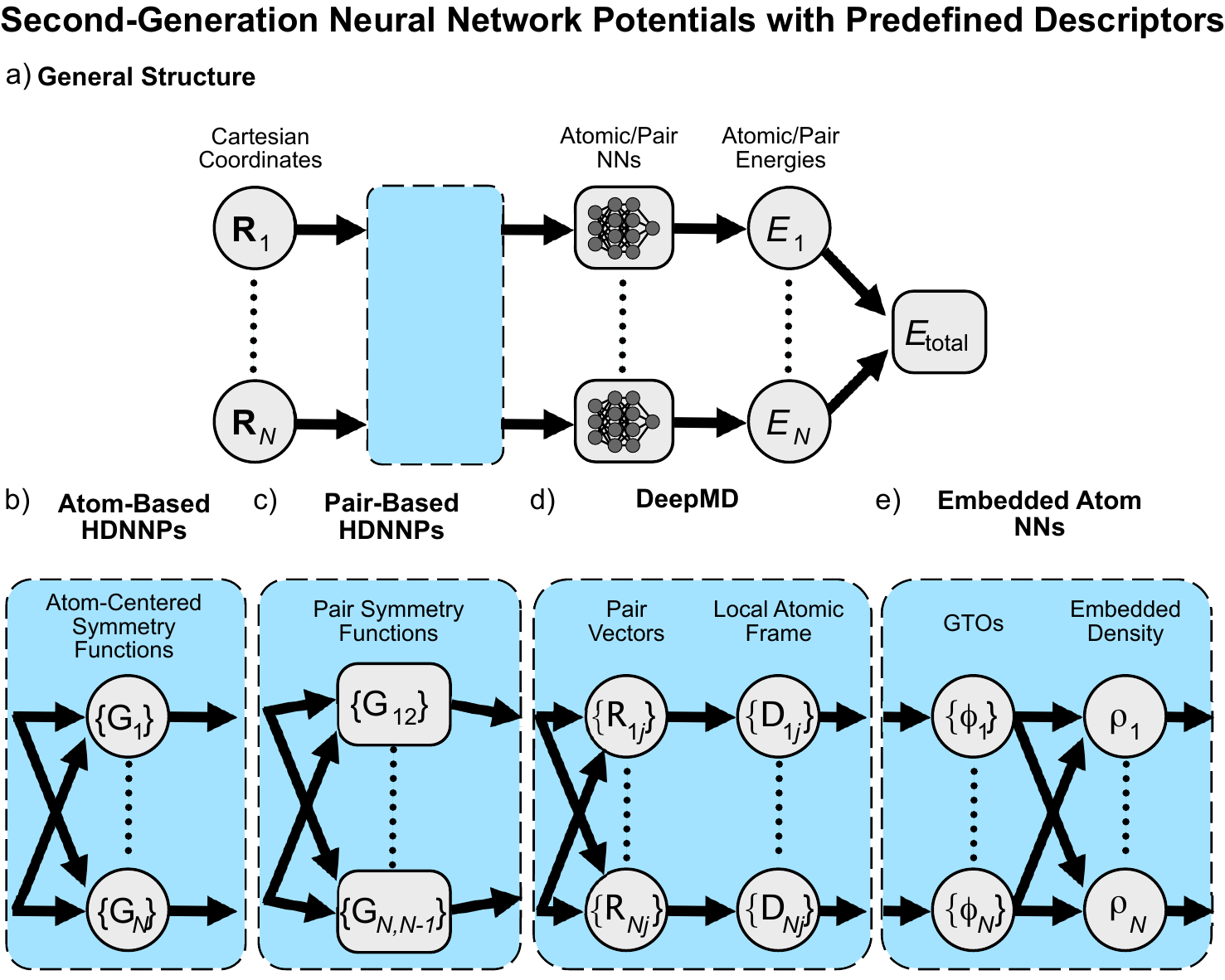}
    \caption{a) General structure of second-generation NNPs with predefined descriptors. The total (short-ranged) energy is the sum of atomic or pair energies provided as outputs of individual atomic or pair neural networks. Starting from the Cartesian coordinates of the atoms $\left\{\mathbf{R}_i\right\}$, different types of transformations to input descriptors characterizing the local atomic environments are performed for b) atom or c) pair-based HDNNPs, d) DeepMD and e) embedded atom NNs. Details of these methods are discussed in Sections~\ref{sec:2ghdnnp}, \ref{sec:pair}, \ref{sec:deepmd} and \ref{sec:eann}, respectively. }
    \label{fig:2GHDNNP}
\end{figure*}
    
\subsection{Predefined Descriptors}
\subsubsection{High-Dimensional Neural Network Potentials} \label{sec:2ghdnnp}

Using Eq.~\ref{eq:2Getot} to construct the energy of the system implies the partitioning of the total energy into atomic contributions, and these contributions can often, to a good approximation, be expressed as a function of the local atomic environment, if the spatial cutoff for describing the environment is chosen large enough to include all energetically relevant atomic interactions. The first MLPs making use of this locality approximation are high-dimensional neural network potentials (HDNNP) proposed by Behler and Parrinello in 2007~\cite{P1174}. In this approach, which for the first time enabled to perform large-scale simulations driven by a MLP, each atomic energy is calculated as the output of an individual atomic neural network (s. Fig.~\ref{fig:2GHDNNP}a,b). To ensure permutation invariance, the architecture and weight parameters of all atomic neural networks for atoms of the same element are constrained to be the same. Alternative, the method can be viewed as using one separate atomic neural network per element, which is evaluated as many times as atoms of the respective element are present in the system. The training of these atom-based second-generation HDNNPs (2G-HDNNPs) is done using energies and forces. The partitioning into atomic energies is carried out automatically in the fitting process allowing the direct use of total energies. Once trained, the HDNNP can then be applied to simulations of systems of variable size by adapting the number of atomic neural networks accordingly, resulting in $O(N)$ scaling of the computational costs of the method. 
\\
The key step for the introduction of HDNNPs has been the development of a novel type of descriptor, atom-centered symmetry functions (ACSF)~\cite{P1174,P2882,P4444}, which for the first time allowed to construct structural fingerprints of the local environments with exact translational, rotational and permutational invariances. Since all structurally equivalent atomic environments give rise to the same symmetry function vector, these invariances are also preserved in the atomic energy outputs of the neural networks. 
\\
The local atomic environments are defined by a cutoff function 
\begin{equation}
         f_{\mathrm{cut}}(R_{i,j}) = 
        \begin{cases}
            0.5\left(\mathrm{cos}\left(\frac{\pi R_{ij}}{R_{c}}\right)+1\right), & \text{if}\ R_{ij} \le R_{\mathrm{c}} \\
            0, & \text{if}\ R_{ij} \ge R_{\mathrm{c}}
        \end{cases}
\end{equation}
which decays smoothly to zero in value and slope at the cutoff radius $R_{\mathrm c}$, and many alternative cutoff functions are available~\cite{P4610,P4551}. $R_{ij}$ is the distance between the reference atom $i$ and any neighboring atom $j$. The cutoff function is a central ingredient of ACSFs and many other similar descriptors, which describe the positions of all neighboring atoms inside the resulting cutoff spheres. Many types of ACSFs have been proposed in the literature to characterize the radial and angular distribution of neighbors~\cite{P2882}. 
The most commonly used radial ACSF has the form
\begin{eqnarray}
    G_{\mathrm{radial},i}=\sum_{j\in R_{\mathrm{c}}} \mathrm{e}^{-\eta(R_{ij}-R_{s})^{2}}\cdot f_{c}(R_{ij})
\end{eqnarray}
with the hyperparameters $\eta$ and $R_s$ defining the shape of a Gaussian sphere around the central atom. Due to the summation over all neighboring atoms the value of $G_{\mathrm{radial},i}$ can be interpreted as a continuous coordination number and a combination of different parameter sets can be used to obtain information about the radial distribution of neighbors. 
In a similar way the angular distribution of neighbors can be probed by the angular function
\begin{equation}
\begin{aligned}
    G_{\mathrm{angular},i}&=2^{1-\zeta}\sum_{j\in R_{\mathrm{c}}}\sum_{k\neq j\in R_{\mathrm{c}}}(1+\lambda \mathrm{cos}\theta_{ijk})^{\zeta}\\& \cdot e^{-\eta(R_{ij}^{2}+R_{ik}^{2}+R_{jk}^{2})} \cdot f_{c}(R_{ij})f_{c}(R_{ik})f_{c}(R_{jk})
\end{aligned}
\end{equation}
with $\theta_{ijk}$ being the angle enclosed by $ij$ and $ik$, and the hyperparameters $\zeta$, $\lambda$ and $\eta$. For multi-element systems, ACSFs for each pair of elements (radial) or each triple of elements (angular) are constructed resulting in a combinatorial increase of the number of symmetry functions if many elements are present.
A detailed discussion of the properties of ACSFs is beyond the scope of this review and can be found in Ref.~\citenum{P2882}. The typical error of energies and forces obtained from HDNNPs is around 1.0 meV and 0.1 eV/$\mathrm{\AA}$, respectively, and in the past decade, HDNNPs have been reported for a variety of systems~\cite{P5128}, from bulk materials via aqueous solutions to interfaces, and also very generally applicable parameterizations of HDNNPs for a broad range of organic molecules, like ANI~\cite{P4945}, which employs a modified angular symmetry function, have been published.

Some attempts have been made in recent years to overcome the combinatorial increase in the number of ACSFs for multi-element systems. In 2018, Gastegger et al. introduced weighted atom-centered symmetry functions (wASCF)~\cite{P5292}. In wACSFs, the ACSFs sharing the same parameters are merged for all element-combinations using element-dependent weighting-prefactors making the input dimensionality of the element-specific atomic neural networks independent of the number of elements in the system.
In a recent study, Liu and Kitchin used wACSFs and proposed to also combine the atomic NNs into one single universal atomic NN with multiple outputs for each chemical species in a system \cite{P5886}. All elements thus share the same connecting weights in the single atomic NN, while the weights connecting the output neurons remain element-specific, which is based on the assumption of a roughly linear relation between a universal feature vector generated by the atomic NN and the element-specific atomic energy, which has been investigated for bulk-elements. A similar method has also been proposed by Profitt and Pearson focusing on molecular systems \cite{P6092}. Finally, the extraction of the relevant information from a large pool of ACSFs can also be done automatically, e.g. by CUR \cite{P5398,P6113} or using convolutional neural networks \cite{P6094}.

\subsubsection{Pair-Based HDNNPs} \label{sec:pair}

The sum over atomic energies in Eq.~\ref{eq:2Getot} is the most frequently used energy expression in high-dimensional MLPs. However, it can equivalently be written as a sum of atom pair energies directly reflecting the atomic interactions as
\\
\begin{eqnarray}
E = \sum_{i=1}^{N_{\mathrm{atoms}}} E_i = \frac{1}{2}\sum_{i=1}^{N_{\mathrm{atoms}}}\sum_{j\ne i}^{N_{\mathrm{atoms}}}E_{ij} \quad .
\end{eqnarray}
\\
In 2012, Behler and coworkers \cite{P3251} suggested to construct the environment-dependent pair energies using pair symmetry functions establishing a new class of descriptors characterizing the combined environments of both atoms in each pair up to a cutoff radius with full translational, rotational and permutational symmetry. The resulting structure of the high-dimensional pair NNP is very similar to conventional atom-based HDNNPs with atom pairs up to a maximum interatomic distance replacing the atoms as central structural entities. For each element combination there is one distinct NN to be trained, which is evaluated as many times as atom pairs of the respective type are present in the system (s. Fig.~\ref{fig:2GHDNNP}c). The method has been shown to be as accurate as the atom-based approach, but because of the larger number of atom pairs compared to atoms the computational costs are increased as there is a larger number of NNs to be evaluated.

An alternative method to express the energy as a sum of bond energies has been proposed by Parkhill in 2017~\cite{P5063}. In this bonds-in-molecule neural network (BIM-NN) a modified version of the bags of bonds descriptor~\cite{P4861} has been employed  using an element-specific distance cutoff to take only covalent bonds into account. The descriptor consists of the bond length as well as information about the directly connected bonds. The resulting pair energies have been interpreted as chemical bond energies. So far the method has only been applied to isolated molecules in vacuum.

A further related approach is AP-Net, which has recently been introduced by Sherrill and coworkers~\citep{P5772}. AP-Net has been developed focusing on non-covalent interactions by representing physically meaningful interaction energies derived from symmetry adapted perturbation theory, which allowed to represent all element combinations by a single neural network for a given type of physical interaction. 

\subsubsection{Deep Potential Molecular Dynamics} \label{sec:deepmd}
 
Deep Potential Molecular Dynamics (DeepMD) developed by E and coworkers~\cite{P5596} is a force-extended version of the Deep Potential method \cite{P5076}, which includes energy training only. Like in HDNNPs, the total energy is written as a sum of environment-dependent atomic energies (Eq.~\ref{eq:2Getot}). For describing the positions of the neighboring atoms, a local atomic frame based on the two closest neighboring atoms is defined. All neighboring atoms are then sorted by element and inverse distance, and for each neighbor $j$ a descriptor vector
\begin{eqnarray}
    \mathbf{D}_{ij}= \{D_{ij}^{0},D_{ij}^{1},D_{ij}^{2},D_{ij}^{3}\}=\{\frac{1}{R_{ij}},\frac{x_{ij}}{R_{ij}^{2}},\frac{y_{ij}}{R_{ij}^{2}},\frac{z_{ij}}{R_{ij}^{2}}\}
\end{eqnarray}
with $x_{ij}$, $y_{ij}$ and $z_{ij}$ being the components of the connecting vector $\mathbf{R}_{ij}$ is defined. Beyond a predefined number of closest neighbors inside the cutoff radius just the first component $1/R_{ij}$ is used corresponding to radial information only. The descriptor vectors of all neighboring atoms are then provided jointly as $\left\{\mathbf{D}_{ij}\right\}$ into atomic neural networks to yield the atomic energies $E_i$ as a function of the geometry of all atom pairs relative to the reference frame (s. Fig.~\ref{fig:2GHDNNP}d). An advantage of DeepMD is the absence of any tunable hyperparameters in the descriptors, which depend only on the atomic positions, reducing the complexity of the potential. Drawbacks are the simple functional forms of the descriptors without explicit many-body information, which therefore needs to be established by the atomic NNs, and the presence of small discontinuities in the forces at the cutoff radius as no smooth cutoff function is applied. The latter limitation has been overcome recently by introducing scalar weight functions in the Deep Potential - Smooth Edition (DeepMD - SE)~\cite{P6088}, in which the atomic neural networks are split into encoding networks determining self-adapted descriptors by mapping the structure to multiple outputs and fitting networks, which are both optimized in the training process.

\subsubsection{Embedded Atom Neural Network Potentials} \label{sec:eann}

In 2019 Zhang, Hu and Jiang proposed the embedded atom neural network potential (EANN) \cite{P5792}, which has been inspired by the embedded atom method (EAM) \cite{P0342}, an empirical potential frequently used in materials science in which each atomic energy is given as an embedding energy arising from the interaction with the density of the surrounding atoms. In EANN, the scalar density value at the position of each atom used in EAM is replaced by a vector of Gaussian-type-orbital (GTO)-based densities, which are used as input for atomic neural networks yielding the atomic energy contributions (s. Fig. \ref{fig:2GHDNNP}e). The expansion coefficients of the GTOs forming angular momentum-specific elements of the density vector are optimized during the training process like in a linear basis set expansion. The density vector of each atom is then used as input vector for the respective atomic NN yielding the atomic energy. Only neighboring atoms up to a cutoff contribute to the density, and a sufficient orbital overlap must be ensured to obtain accurate potentials. A transformation to an angular basis makes the method computationally very efficient, as three-body descriptors do not need to be computed explicitly.

\subsection{Learnable Descriptors} \label{ref:2GMP}
\subsubsection{Overview: Message Passing Neural Networks}

All NNPs discussed so far have in common that the descriptor functions for the atomic structure are predefined and often contain some hyperparameters, which may be manually selected or adapted during the training process. The purpose of these descriptors is to provide local structural fingerprints of the atomic environments as input for atomic neural networks. Thus, the only necessary information is the atomic structure, and a vector of function values needs to be calculated to characterize each environment as  uniquely as possible with as few as possible functions. 

In the context of ML applications in molecular modelling, the idea of replacing predefined ``static'' descriptors by learned ``dynamic'' ones using structural information was first introduced by Duvenaud and coworkers \cite{P5307}. Inspired by the extended-connectivity circular fingerprints (ECFP) \cite{P6061}, they treated molecules as graph networks and employed a convolutional layer to process these graphical networks to predict their final feature vectors, i.e., the descriptors. Although they did not exploit these methods to construct PESs, this work paved the way for a new class of MLP, for which Gilmer et al. in 2017 coined the name ``message passing neural networks'' (MPNN)~\cite{P5368}. All MPNN methods have in common that predefined descriptors are replaced by automatically determined descriptors learned from the geometric structure. 

In MPNNs, a molecule is considered as a three-dimensional graph consisting of nodes (vertices) and connections (edges), which are associated with the atoms defined by their nuclear charges $\{Z_{i}\}$ and bonds, or more generally interatomic distances $\{R_{ij}\}$, respectively. Each atom is described by a feature vector, which is iteratively updated in an interaction block in a message passing phase using message functions processing information about neighboring atoms, like distances and their respective current feature vectors. The various flavors of MPNNs differ in the types of message functions that iteratively embed atoms into their chemical environments yielding the messages interacting with the atomic feature vectors, the atomic update functions (vertex update functions) describing these interactions, and the distance of neighbors included.
Once the message passing phase has been completed after a predefined number of steps, the resulting atomic feature vectors, which now contain the information about the atom in its environment, are used in a readout phase to predict the target property. In the context of NNPs, the atomic feature vectors are typically passed to atomic neural networks in this step to yield atomic energy contributions summing up to the total energy.

\begin{figure*}
\centering
\includegraphics[width=14cm]{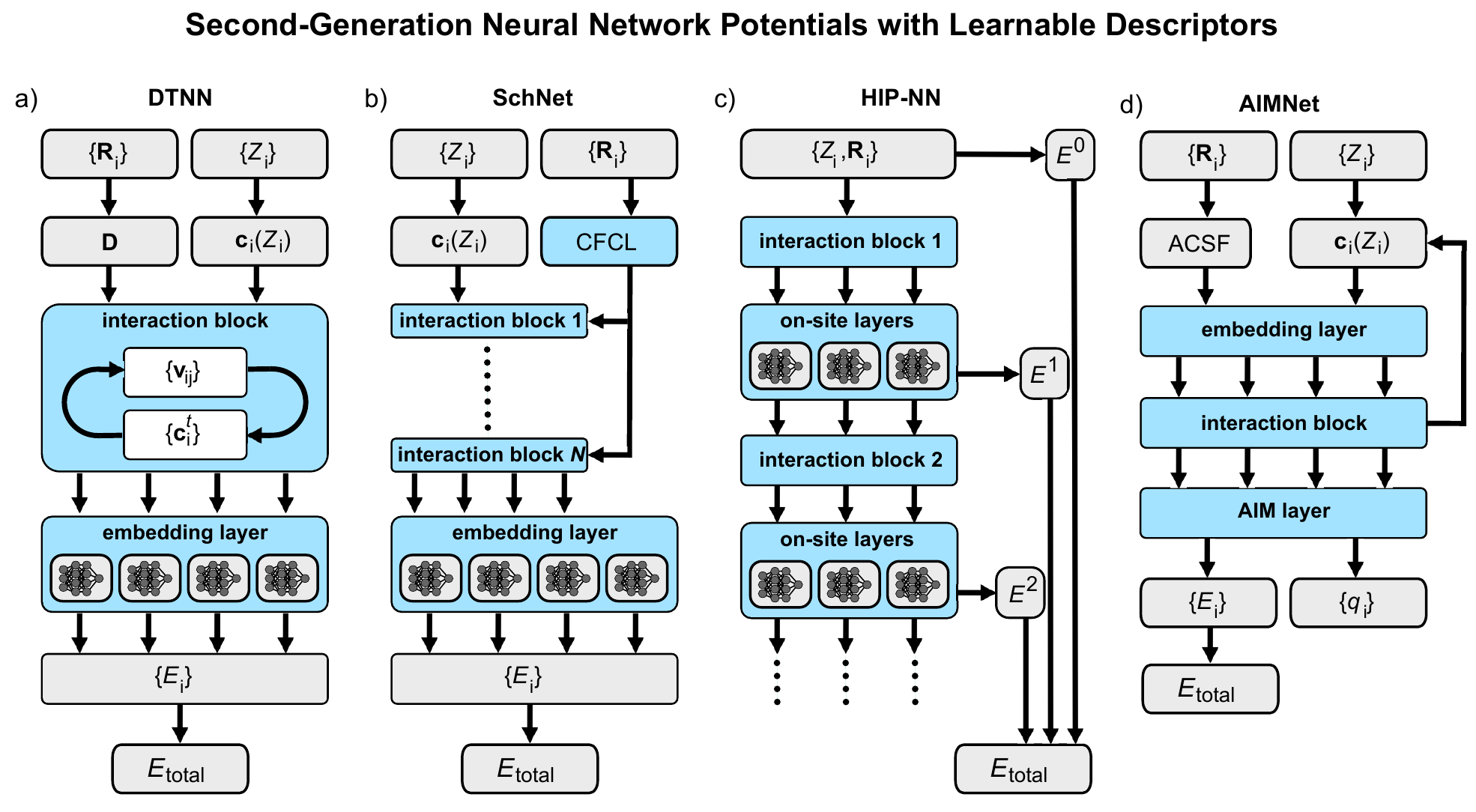}
\caption{Schematic structure of the message passing neural networks DTNN~\cite{P4937}, SchNet~\cite{P5887}, HIP-NN~\cite{P5888}, AIMNet~\cite{P5817} and PhysNet~\cite{P5577} discussed in this review in Sections \ref{sec:dtnn}, \ref{sec:schnet}, \ref{sec:hipnn}, \ref{sec:aimnet} and \ref{sec:3g}, respectively.}
    \label{fig:mpnn}
\end{figure*}

\subsubsection{Deep Tensor Neural Networks (DTNN)} \label{sec:dtnn}

The first MLP employing message passing neural networks has been the deep tensor neural network (DTNN) proposed by Sch\"utt and coworkers in 2017~\cite{P4937}, which has been designed to describe the properties of small organic molecules. Starting from the nuclear charges and a complete distance matrix $\mathbf{D}$, the pairwise interatomic distances $D_{ij}$ are first expanded in a Gaussian basis yielding a coefficient vector $\mathbf{d}_{ij}$ for each atomic pair (s. Fig.~\ref{fig:mpnn}a). Further, each atom $i$ is represented by a multidimensional feature vector $\mathbf{c}_{i}$ starting with an initial random nuclear charge-specific vector $\mathbf{c}_{i}({Z_i})$. The atomic feature vectors are then iteratively refined in an interaction block using message vectors $\mathbf{v}_{ij}$ reflecting the interactions of atom $i$ with all other atoms $j$ by 
\begin{equation}
    \mathbf{c_{i}}^{t+1}=\mathbf{c_{i}}^{t}+\sum_{j\neq i}^{N_{\mathrm{atoms}}}\mathbf{v}_{ij} \quad.
\end{equation}
In total, $T=3$ of these global refinement steps are carried out in DTNN~\cite{P4937}, after which the accuracy has been found to saturate for the investigated systems. The $\mathbf{v}_{ij}$ contain the non-linear coupling of the neighboring atomic feature vectors and the corresponding interatomic distance vectors and are computed in a tensor layer,
\begin{equation}
    \mathbf{v_{ij}}=\mathrm{tanh}\left[\mathbf{W}^{\mathrm{fc}}\left(\left(\mathbf{W}^{\mathrm{cf}}\mathbf{c}_j+\mathbf{b}^{\mathrm{f}_1}\right)\circ\left(\mathbf{W}^{\mathrm{df}}\mathbf{d}_{ij}+\mathbf{b}^{\mathrm{f}_2}\right)\right)\right] \quad,
\end{equation}
where $\circ$ is the element-wise multiplication and $\mathbf{W}^{\mathrm{fc}}$, $\mathbf{W}^{\mathrm{cf}}$, $\mathbf{W}^{\mathrm{df}}$, $\mathbf{b}^{\mathrm{f}_1}$ and $\mathbf{b}^{\mathrm{f}_2}$ are weight and bias matrices to be trained. Consequently, the atomic feature vectors start from the state of free atoms and are then iteratively refined using information about the distances and feature vectors of all other atoms.
Once this refinement step through convolutional mapping is complete, the feature vectors are provided to an embedding layer containing atomic NNs yielding atomic energies as output entering Eq.~\ref{eq:2Getot}.

The accuracy of DTNNs has been demonstrated in compositional space for the atomization energies of optimized molecules, in configurational space for the isomers of composition C$_7$O$_2$H$_{10}$ and in conformational space for molecular dynamics trajectories of small molecules in vacuum~\cite{P4937}.
Due to the use of a complete set of interatomic distances DTNNs represent a global method applicable to small molecules, but in principle a spatial cutoff could be introduced for the message vectors to extend the method to a second-generation potential.

\subsubsection{SchNet} \label{sec:schnet}

Briefly after the introduction of DTNNs, Sch\"utt et al. proposed an improved MPNN called SchNet~\cite{P5887,P5366} (s. Fig.~\ref{fig:mpnn}b), which shares many properties with the original DTNN approach.
In SchNet, so-called continuous-filter convolutional layers (CFCL) are used to represent the atomic environments, which are continuous generalizations of the discrete convolutional networks typically used in image analysis applications, since in contrast to pixel grids in images atomic positions are continuous quantities. In this way the coefficient vector of the Gaussian expansion of the interatomic distances can be replaced by the atomic positions. After the atomic feature vectors have been randomly initialized based on their species, they are distributed to a set of $N$ interaction blocks in which CFCLs act on each local atomic neighborhood through a message passing phase where they iteratively refine structural representations by processing atomic feature vectors. For the interatomic distances, a cutoff is applied making the method applicable to systems with large number of atoms including periodic systems like crystals. Like in DTNN the parameters of the atomic neural network layers are shared for all atoms, and different elements are distinguished based on their features. After the refinement of the atomic feature vectors in the iteraction blocks, they are used as input of atomic neural networks yielding the atomic energy contributions, which are added to obtain the total energy of the system.

\subsubsection{Hierarchically Interacting  Particle  Neural  Network} \label{sec:hipnn}

A NNP combining the concepts of MPNNs and conventional many-body expansions is the ``Hierarchically Interacting Particle Neural Network'' (HIP-NN) proposed by Lubbers, Smith and Barros in 2018~\cite{P5888}. Going beyond Eq.~\ref{eq:2Getot}, the atomic energies $E_i$ are further decomposed into contributions of order $n$,
\begin{eqnarray}
E_i=\sum_{n=0}^{N_{\mathrm{body}}}E_i^n \quad , \label{eq:HIPNN}
\end{eqnarray}
which are learned simultaneously and consistently by a single hierarchical NN consisting of many layers  and intermediate interaction blocks (s. Fig.~\ref{fig:mpnn}c). The basis of the structural description are the nuclear charges and those pairwise interatomic distances, which are below a predefined cutoff distance. The term $E^0=\sum E_i^0$ is computed to minimize the least squares error of the training data using nuclear charge information only. The first interaction block corresponds to the first message passing step in that information between atoms within a cutoff radius is exchanged. The atomic feature vectors are then learned through a sequence of independent atomic on-site layers the last of which corresponds to a  linear regression on the learned atomic descriptor vectors and yielding  $E^1=\sum E_i^1$. Then, the second interaction block represents another message passing step, followed by atomic on-site layers to yield features to compute $E^2$ and so forth. In HIP-NN, this sequence is terminated after the $n=2$ term.
The total energy is then given according to Eq.~\ref{eq:HIPNN} as the sum of all computed terms.
As it is assumed that the energy contributions decay with increasing order, the magnitude of the different terms can be used to estimate the uncertainty of the HIP-NN prediction.

\subsubsection{AIMNet} \label{sec:aimnet}

Another NNP with learnable descriptors called AIMNet has been proposed by Isayev and coworkers in 2019~\cite{P5817} (s. Fig.~\ref{fig:mpnn}d). Starting from the atomic positions and nuclear charges, the  atomic feature vectors are constructed by combining two components, which are the element-independent geometric description by atomic environment vectors consisting of modified ACSFs of the ANI HDNNP~\cite{P4945} employing a cutoff, and the initial feature vectors containing the information about the nuclear charges. Therefore, like in DTNN and SchNet, the dimensionality of the resulting atomic feature vectors does not depend on the number of chemical elements in the system. In addition to the atomic feature vectors, also atom-pair feature vectors are used to describe the bonds in the system.

In contrast to other MPNNs, in AIMNet an embedding layer can be used directly after the first feature vectors have been constructed, because the latter already contain right from the start information about the atomic environments via the ACSFs and do not rely on a preparing message passing step.
Like e.g. in SchNet, the effective cutoff of atomic interactions can be increased by multiple repeated updated steps of the feature vectors in interaction blocks. Finally, an atom-in-molecule (AIM) layer then allows to compute the atomic properties.

Apart from the environment-dependent atomic energies yielding the total energy, other properties like atomic partial charges can be provided by AIMNet, which can be redistributed within the interaction range covered by the sequence of message passing steps. Typically, in AIMNet the AIM layer is calculated three times corresponding to two message passing updates of the feature vectors.
Long-range interactions like electrostatics are not explicitly included beyond the distance covered by these message passing steps. A recent extension, AIMNet-ME~\cite{P6037}, can also be trained to systems in different global charge states by an additional message-passing stack for the spins and charges in the system.

\section{The Third Generation: Long-Range Interactions} \label{sec:3g}

Second-generation NNPs, as well as second-generation MLPs in general, have been applied very successfully to a wide range of systems, and they represent today's workhorse methods in machine learning-based atomistic simulations. Still, the underlying locality approximation and the neglect of interactions beyond the cutoff radius can be expected to result in notable errors for systems, in which long-range interactions are important. The most prominent example for such long-range interactions is electrostatics, but also comparably weak dispersion interactions can amount to decisive energy contributions for large systems \cite{P4403}.
\\
A basic prerequisite for considering long-range electrostatic interactions is the availability of atomic partial charges. Starting with the pioneering work of Popelier and coworkers~\cite{P2391,P2211}, who explored the capabilities of NNs and other machine learning methods to represent electrostatic mono- and multipoles for the improvement of electrostatics in classical force fields, many groups have proposed to represent atomic properties like charges and multipoles by machine learning \cite{P5310,P5205,P5927,P5933,cuevas2021machine,wang2019graph,wang2021deepatomiccharge}. However, to date only a few NNPs contain long-range electrostatic interactions by explicitly computing the Coulomb interaction without truncation. These potentials define the third generation of NNPs.
\\
The first MLP of the third generation has been the 3G-HDNNP introduced in 2011 \cite{P2962,P3132}. In this method, which is an extension of the second-generation HDNNPs proposed by Behler and Parrinello (s. Sec.~\ref{sec:2ghdnnp}), in addition to the atomic NNs yielding the short-range atomic energies, a second set of atomic NNs is introduced to predict environment-dependent atomic charges, which are then employed in an Ewald summation~\cite{P0238} to compute the electrostatic energy $E_{\mathrm{elec}}$. Often the same ACSF vector $\mathbf{G}_i$ for describing the atomic environment is used. The total energy expression of this method is given by
\begin{equation}
\begin{aligned}
E_{\mathrm{total}}&=E_{\mathrm{short}}+E_{\mathrm{elec}}\\ &=\sum_{i=1}^{N_{\mathrm{atoms}}}E_{i}\left(\{\mathbf{G}_{i}\}\right)+\sum_{i> j}^{N_{\mathrm{atoms}}}\frac{q_{i}(\{\mathbf{G}_{i}\})q_{j}(\{\mathbf{G}_{j}\})}{R_{ij}}
\end{aligned}
\end{equation}
Since the short-range atomic energies can describe in principle all types of atomic interactions including electrostatics up to the cutoff radius, double counting of energy terms has to be avoided by a sequential training process.  First, the atomic neural network representing the charges are trained using reference charges from electronic structure calculations. Then, the electrostatic energies and forces are computed and removed from the reference energies and charges to obtain the target properties needed for training the short-range energies. This two-step procedure is also required since the computation of electrostatic forces resulting from environment-dependent charges required the the derivatives of the atomic charges with respect to the atomic positions, which can be provided by the charge NNs but are not accessible in electronic structure calculations.
Another third-generation HDNNP containing long-range electrostatics and also dispersion interactions using Grimmes D2 method~\cite{grimme2006semiempirical} is the Tensormol approach introduced in 2018 by Yao et al.~\cite{P5313}. In this method, the charges are trained by NNs to reproduce reference dipole moments obtained in electronic structure calculations. 

In addition to HDNNPs relying on predefined descriptors, also MPNNs (s. Section \ref{ref:2GMP}) including long-range interactions have been reported. The most important example is PhysNet introduced by Unke and Meuwly in 2019~\cite{P5577}. This method is closely related to SchNet with two important modifications: First, the information flow in the interaction block is enhanced by pre-activation residual layers that enable a higher level of expressiveness and distance-based attention masks are used as messenger functions to increase the efficiency of the learning cycle. The second major component is the explicit treatment of long-range interactions and electrostatics. In PhysNet, atomic coefficient vectors pass through one single refinement box that contains many interaction blocks and are then fed into atomic neural networks to predict the atomic energies and charges, which are in turn used to compute the long-range electrostatic energy. An important feature of this approach is the simultaneous prediction of both, atomic energies and charges, by the same NN increasing the computational efficiency of the method.

In spite of these advances, third-generation NNPs are rarely used, as in many condensed systems long-range electrostatic interactions are efficiently screened contributing only very little to the atomic interactions beyond the typical cutoff radii of 6-10~\AA{}. In addition, the explicit computation of the Ewald sum, as well as the introduction of additional NNs in several methods, substantially increases the computational costs with very small improvements in accuracy.

\section{The Fourth Generation: Non-Local Interactions}

\subsection{Overview: Non-Local Dependencies}

Even though long-range electrostatic interactions are included in third-generation NNPs, the underlying charges are assumed to be local in that they only depend on the positions of the neighboring atoms up to a cutoff radius. While this is a reasonable assumption for many systems, the locality approximation breaks down for systems in which the atomic partial charges depend on  structural features outside the atomic cutoff spheres. These can be distant functional groups in molecules or even ion substitution, doping and defects in solid materials. Another typical situation involving non-local modifications of the electronic structure is a change in the global charge state of the system, e.g. by ionization, protonation or deprotonation. In these situations the first three generations of NNPs, which implicitly assume a single fixed total charge, are unable to represent the PES reliably, since it is impossible to represent the atomic partial charges as a function of the local environment only.

\begin{figure*}
    \centering
    \includegraphics[width=14cm]{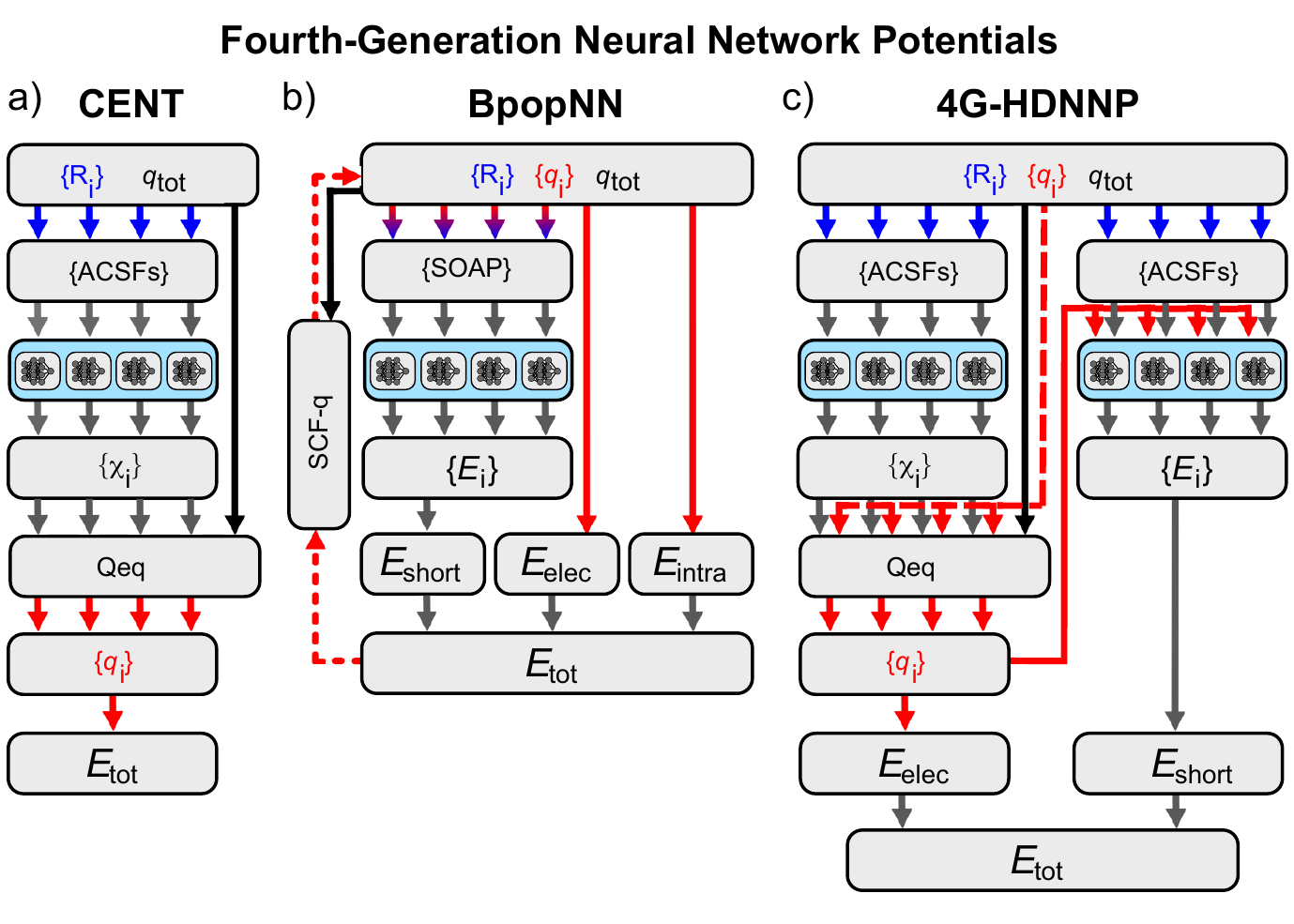}
    \caption{Schematic structure of fourth-generation neural network potentials. In the CENT method~\cite{P4419} the atomic electronegativities $\chi_i$ are expressed by atomic NNs as a function of the local atomic environments described by atom-centered symmetry functions (ACSF). They are then used in a charge equilibration (Qeq) step to provide globally dependent charges used to minimize the energy in Eq.~\ref{eq:min_Qeq}. In the Becke population neural network (BpopNN)~\cite{P5859} a charge-dependent SOAP descriptor is used to determine the atomic energies. Atomic charges are also used to compute electrostatic and intra-atomic energy terms. When used in predictions, the atomic charges are determined self-consistently (SCF-q) by minimizing the total energy as indicated by the dotted line. In Fourth-generation high-dimensional neural network potentials (4G-HDNNP)~\cite{P5932} the atomic charges are determined in a charge equilibration step like in CENT, but the training is done using reference charges as indicated by the dashed line. Apart from the calculation of the electrostatic energy, the charges are also used as additional non-local input information for the atomic energies $E_i$ provided by a second set of atomic neural networks to yield the short-range energy. The total energy is the sum of the short-range and the electrostatic energy.}
    \label{fig:4G_potentials}
\end{figure*}

Potentials, which are able to capture these non-local or even global dependencies for high-dimensional systems define fourth generation NNPs. We note that the terminology in the literature is not consistent because often there is no clear distinction between long-range interactions and non-local interactions. In this review, long-range interactions refer to electrostatics, and also van der Waals interactions, which are not truncated, but depend on local properties, like charges, and hence can be described by third-generation NNPs. Non-local interactions arise from long-range or even global dependencies in the electronic structure and require fourth-generation NNPs for a qualitatively correct description. 
In recent years, a few NNPs of the fourth generation have been proposed (cf. Fig. ~\ref{fig:4G_potentials}), which will be discussed in the following Sections.

\subsection{Methods} 
\subsubsection{Charge Equilibration Neural Network Technique} \label{sec:cent}

The first MLP of the fourth generation has been the charge equilibration via neural network technique (CENT) proposed by Ghasemi et al. in 2015~\citep{P4419}. The central idea of CENT is to employ a charge equilibration step~\citep{P1448}, which is also used in several advanced force fields and allows the redistribution of the electrons over the whole system to minimize the electrostatic energy. The total energy of CENT is based on a second order Taylor expansion in the atomic partial charges $q_i$ and the classical electrostatic energy of the charge density and is given by
\begin{eqnarray}
E_{\mathrm{CENT}}&=&
\sum_{i=1}^{N_{\mathrm{atoms}}} \left(E_{i,0} + \chi_{i} q_{i} + \frac{1}{2}
\left(\eta_{i}+\frac{2 \gamma_{ii}}{\sqrt{\pi}}\right) q_i^2 \right) \nonumber \\ 
&&+ \sum_{i>j}^{N_{\mathrm{atoms}}} q_{i} q_{j} \frac{\mathrm{erf}(\gamma_{ij} R_{ij})}{R_{ij}} \quad.  \label{eq:min_Qeq} 
\end{eqnarray}
The $E_{i,0}$ are the energies of the free atoms, the coefficients of the linear terms are the atomic electronegativities $\chi_{i}$, and the $\eta_i$ represent the element-dependent hardness values. The atomic electronegativities are expressed by individual atomic NNs as a function of the local atomic environments described by ACSFs (see Fig.~\ref{fig:4G_potentials}a). The parameters $\gamma_{ij}$ depend on the width of the atomic Gaussian charge densities. 

The minimization of this energy expression results in a set of linear equations that in addition contains a constraint for the total charge of the system. This needs to be solved to yield the equilibrated atomic partial charges in the system. The weights of the electronegativity NNs are optimized during the training process such that the error of $E_{\mathrm {CENT}}$ with respect to DFT reference energies is minimized. Like all fourth-generation MLPs, the CENT method allows to construct potentials, which are simultaneously applicable to several global charge states, and due to the charge-based energy expression Eq.~\ref{eq:min_Qeq} the method works best for systems with predominantly ionic bonding.

\subsubsection{Becke Population Neural Networks}

Another fourth-generation NNP is the Becke population neural network (BpopNN) introduced in 2020 by Xie, Persson and Small~\citep{P5859}. In this method, for the chosen global charge of the system the atomic populations, i.e., the partial charges, are adapted self-consistently in a process called SCF-q (see Figure~\ref{fig:4G_potentials}b). Starting from an initial guess of the charge distribution, the total energy of the system is minimized with respect to the atomic populations and calculated as
\begin{equation}
    E_{\mathrm{BpopNN}}=\sum_{i=1}^{N_{\mathrm{atoms}}}E_{i}(\mathbf{D}_{i})+E_{\mathrm{intra}}+E_{\mathrm{elec}} \quad , \label{eq:ebpopnn}
\end{equation}
where the $E_{i}$ represent the energy contributions of atoms $i$ determined by atomic neural networks as a function of both, the local atomic environments as well as the atomic populations, which are provided as input in form of modified SOAP~\cite{P3885} descriptors $\mathbf{D}_{i}$. The intra-atomic energy $E_{\mathrm{intra}}$ is an element-specific quadratic function of the atomic partial charge. 
The electrostatic energy $E_{\mathrm{elec}}$ contains a damping function at very short interatomic distance facilitating the representation of the atomic short-range energies. 

In the training stage, the reference atomic populations, total energies and atomic forces are obtained from constrained DFT calculations~\cite{P3083} for various charge distributions and states. Then, the atomic NN weights and element-pair dependent parameters $\kappa(z_{i},z_{j})$ are adjusted such that the errors of the total energy, the atomic forces and the population gradients is minimized with respect to the DFT data. 

\subsubsection{Fourth-Generation High-Dimensional Neural Network Potentials}

Very recently, a fourth generation of HDNNPs (4G-HDNNP) has been proposed~\citep{P5932,P5977} combining the advantages of the CENT method (s. Sec. \ref{sec:cent}), i.e., the ability to describe non-local charge transfer, global dependencies in the electronic structure and also the resulting electrostatic interactions, and of second-generation HDNNPs (s. Sec. \ref{sec:2ghdnnp}), which provide a very accurate description of local bonding. The schematic structure of a 4G-HDNNP is shown in Fig.~\ref{fig:4G_potentials}c. Like in CENT, the electrostatic energy is computed from charges obtained in a charge equilibration process based on environment-dependent electronegativities, which are represented by atomic NNs. In contrast to CENT, in the training process the electronegativities are not adjusted to yield charges minimizing the total energy, but the NN weights are determined to reproduce a set of reference Hirshfeld charges obtained in DFT calculations (dashed red line in Fig.~\ref{fig:4G_potentials}c). The long-range energy is then calculated either using Coulomb's law or, in case of periodic systems, using an Ewald sum~\cite{P0238}.

The second energy contribution, the short-range energy, is given as a sum of atomic energies like in second-generation HDNNPs, but in addition the atomic partial charges of the respective atoms are used as additional input neurons to provide information about the local electronic structure. For the same reasons as in third-generation HDNNPs, the short-range atomic NNs are trained in a second step after the electrostatic NNs, which is also required since the atomic charges used as additional input for the short-range atomic NNs need to be available.

The resulting 4G-HDNNPs are very generally applicable to many types of systems, from organic molecules to ionic solids~\cite{P5932}. Like all fourth-generation potentials, they can be constructed to simultaneously describe different global charge states of a given system.

\section{Discussion and Outlook}

Starting with the applicability to high-dimensional systems in second-generation potentials, MLPs for atomistic simulations have attracted significant attention resulting in a rapid development of this field, which is now very advanced but has not yet reached full maturity. While initially applications in materials science have been the main driving force for these developments \cite{P1174}, MLPs are now explored and applied to many different fields, and a lot of effort is nowadays spent, e.g., in the construction of broadly applicable potentials for a wide range of organic molecules \cite{P4945,P5818,P4937,P5577}.

The development of MLPs is full of challenges, some of which have been solved while others require further research. The development of descriptors for high-dimensional systems considering all symmetries and invariances exactly, which in the initial years has been a formidable challenge, can now be considered as solved, and there are two alternative approaches based on predefined or learnable descriptors. Both are equally suited for high-quality potentials, but since the first message-passing networks have been introduced only a few years ago, most applications published to date rely on predefined descriptors like ACSFs~\cite{P2882}, SOAP~\cite{P3885} and many others~\cite{P5803}. Still, a yet unsolved problem is the combinatorial growth in the number of predefined descriptors with increasing number of elements in the system. First promising steps, like combined descriptors~\cite{P5292} or even combined atomic NNs~\cite{P5886,P6092} have been proposed, and also many MPNNs make use of combined feature vectors~\cite{P4937,P5366,P5817}, but the generality of these methods remains to be explored.

Another challenge, which is addressed by many groups, is the efficient construction of the reference data sets, which is of crucial importance for obtaining reliable potentials. Since the computationally most expensive part for constructing NNPs is the generation of the reference data  by accurate electronic structure calculations, the data sets should be as small as possible, while at the same time they should cover a broad range of structures to obtain transferable potentials. Nowadays, the concept of active learning, which has been known in the ML field for a long time~\cite{P5900} and has first been used by Artrith and Behler for the construction of a HDNNP for copper~\cite{P3114}, is now broadly applied to establish an almost automatic framework for constructing reference data. The central idea of active learning is to identify important missing structures without an explicit comparison to costly electronic structure calculations, but to use an estimate for the uncertainty of a prediction to select new data points. This can, e.g., be done by comparing the energies or forces obtained from different NNPs trained to the same data set, and many different variants for the iterative construction of the reference data by active learning have been proposed to date~\cite{P4939,P6042,P5782,P6023,P5399,P5842,P6103,P5738}.

Another approach to improve the quality of MLPs with only a moderate amount of high-level data is $\Delta$-learning~\cite{P5959,P4513,P6106}. Here, assuming that the difference between two PESs obtained with different electronic structure methods is smooth and thus relatively straightforward to represent by ML, a baseline potential is obtained using an approximate method, and the difference to the high-level method is learned by the NNP.

Apart from the optimized selection of the underlying data~\cite{P5398}, another main strategy to generate more accurate and transferable potentials is to include the underlying physical laws. An example is the inclusion of long-range electrostatic interactions in third and fourth-generation NNPs. While most fourth-generation NNPs employ predefined descriptors, also novel types of message passing methods are just emerging aiming to describe non-local effects~\cite{P5927,P5933,P6056}. Apart from electrostatics,  also dispersion interactions, which are weak but can be important in large systems, have been included beyond the local atomic environments in NNPs~\cite{P5313,P5577}. More recently, information beyond charges such as atomic spin moments~\citep{P6057,P6056} for magnetic systems, and external electric field~\citep{P6108} to study molecular spectra have also been considered.

A new direction of the field is to combine NNPs with information from electronic structure methods of different levels of complexity. There are several examples like the combination of NNs with density-functional tight binding~\cite{P5605} or H\"uckel theory~\cite{P6046}, and the representation of symmetry-adapted atomic-orbital features~\citep{P6044} by message passing techniques.  Furthermore, highly accurate molecular electronic wavefunctions can be learned by NNs, which include the representation of electrons in the molecule~\citep{P6081,P6080}. This trend of including or generating more information at the electronic structure or quantum mechanical level can be expected to continue in the years to come.

In summary, NNPs have become an indispensable tool for atomistic simulations in many fields of chemistry, molecular biology and materials science. They have been proven to be particularly useful for large-scale molecular dynamics simulations in extensive sampling problems, which require long simulations of extended systems. Since they are able to describe all types of bonding with the same level of accuracy, they are even applicable to systems, which are difficult to describe by conventional types of potentials. Still, a main limitation of NNPs, and MLPs in general, is the  restricted transferability beyond the underlying training set, but it can be anticipated that future generations of potentials will allow to overcome this limitation by the inclusion of physical knowledge and laws while maintaining the general applicability, which is a major advantage of the family of MLPs.

\section*{DISCLOSURE STATEMENT}
The authors are not aware of any affiliations, memberships, funding, or financial holdings that
might be perceived as affecting the objectivity of this review. 

\section*{ACKNOWLEDGMENTS}
Financial support by the Deutsche Forschungsgemeinschaft (DFG) is gratefully acknowledged (BE3264/13-1, project number 411538199).

\bibliography{literature}

\end{document}